 \documentclass[final,3p,times]{elsarticle}

\usepackage{amssymb}
\usepackage{amsthm}

\usepackage{makeidx}  
\usepackage{algorithm}
\usepackage[noend]{algorithmic}
\usepackage{qtree}
\usepackage[table]{xcolor}
\usepackage{graphicx}
\usepackage{graphics}
\usepackage{amsmath}
\usepackage{amsfonts}  
\usepackage{parskip}

\biboptions{sort&compress}

\journal{Journal of Computer and System Sciences}
\begin{document}
\begin{frontmatter}



\newtheorem{definition}{Definition}
\newtheorem{theorem}{Theorem}[section]
\newtheorem{lemma}[theorem]{Lemma}
\newtheorem{proposition}[theorem]{Proposition}
\newtheorem{corollary}[theorem]{Corollary}

\hyphenation{}

\title{Mixing local and global information for community detection in large networks}


\author[1]{Pasquale De Meo \ }
\address[1]{University of Messina, Department of Mathematics and Informatics. V.le F. Stagno D'Alcontres 31, I-98166 Messina, Italy}
\ead{pdemeo@unime.it}
\author[2,1]{Emilio Ferrara \ }
\address[2]{Center for Complex Networks and Systems Research, School of Informatics and Computing.\\
Indiana University Bloomington, 919 E. 10th St., Bloomington, IN 47408, USA}
\ead{ferrarae@indiana.edu}
\cortext[cor1]{Corresponding author}
\author[1]{Giacomo Fiumara \ }
\ead{gfiumara@unime.it}
\author[1,3]{Alessandro Provetti \ \corref{cor1}}
\address[3]{Oxford-Man Institute of Quantitative Finance, University of Oxford. Oxford. OX2 6ED, UK}
\ead{ale@unime.it}

\address{}

\begin{abstract}
The problem of clustering large complex networks plays a key role in several scientific fields ranging from Biology to Sociology and Computer Science.
Many approaches to clustering complex networks are based on the idea of maximizing a {\em network modularity} function.
Some of these approaches can be classified as {\em global} because they exploit knowledge about the whole network topology to find clusters.
Other approaches, instead, can be interpreted as {\em local} because they require only a partial knowledge of the network topology, e.g., the neighbors of a vertex.
Global approaches are able to achieve high values of modularity but they do not scale well on large networks and, therefore, they cannot be applied to analyze on-line social networks like Facebook or YouTube.
In contrast, local approaches are fast and scale up to large, real-life networks, at the cost of poorer results than those achieved by local methods.

In this article we propose a {\em glocal} method to maximizing modularity, i.e., our method uses information at the global level, yet its scalability on large networks is comparable to that of local methods.

The proposed method is called COmplex Network CLUster DEtection (or, shortly, CONCLUDE.)
It works in two stages: in the first stage it uses an information-propagation model, based on random and non-backtracking walks of finite length, to compute the {\em importance} of each edge in keeping the network connected (called {\em edge centrality.)}
Then, edge centrality is used to map network vertices onto points of an Euclidean space and to compute distances between all pairs of connected vertices.
In the second stage, CONCLUDE uses the distances computed in the first stage to partition the network into clusters.

CONCLUDE is computationally efficient since in the average case its cost is roughly linear in the number of edges of the network.
Testing on diverse benchmark datasets shows good results, both in times and quality of the clustering; that is the case for synthetic networks with well-defined clusters as well as for real-world network instances.
\end{abstract}


\end{frontmatter}

\section{Introduction} \label{sec:introduction}
In a network, the term {\em community structure} indicates the presence of groups of vertices called {\em communities} or {\em clusters} such that there is a large number of edges connecting vertices inside the same community and few edges linking vertices located in different clusters \cite{fortunato2010community}. 
For a given network, represented by a graph $G = \langle V, E \rangle$ where $V$ is the set of vertices and $E$ the set of edges, the {\em community detection problem} consists of finding a partition of the vertices in $V$ of the form $\mathcal{C} = \{ \mathcal{C}_1, \ldots, \mathcal{C}_{q} \}$ such that each $\mathcal{C}_i$, $1 \leq i \leq q$ exhibits the community structure described above.


The detection of clusters in large graphs is becoming a central research problem in a variety of different areas including VLSI design, parallel computing, computer vision and social network analysis \cite{fortunato2010community}.
Such an interest depends on the fact that many real-world systems consist of separated modules that interact with each other. 
If modules are clearly identified, it is then possible to understand the role each module plays in the overall behavioral dynamics of the system, and to study how potential changes in the structure and functions of a module would impact the overall system. 
As an example, in the biological domain community detection algorithms have been deployed to clarify the functioning of metabolic networks \cite{GuAm05} or to understand how some proteins interact in small groups (or subsystems) called 'modules,' or forming so-called \emph{complexes} \cite{von2003genome}. 
In Computer Science and Sociology, community detection algorithms have been exploited to understand the social structures arising from the interactions of single individuals; 
this has relevant practical applications, for instance in customer segmentation.


Early methods for finding communities were rooted in graph theory, e.g., the principle of maximum flow and minimum cut \cite{fortunato2010community}. 
The main limitation of these methods is their high computational cost, which restricts their applicability to toy models or small instances (e.g., samples) of real-world networks.

Recently, attention is increasingly paid to {\em spectral clustering methods} \cite{kempe2008decentralized,ng2001spectral,shi2000normalized,weiss1999segmentation}, which strive to optimize suitable cost functions.
Empirical analysis provides evidence that spectral clustering methods are able to achieve excellent performance in some domains,  e.g., image segmentation.
Most of the spectral clustering methods are parametrized to the desired number of clusters, denoted by an integer \textit{k.}
In many cases, spectral clustering methods tend to produce clusters of (almost) equal size. 
However, in a broad range of application domains, such features have been considered as serious drawbacks, especially since often there is no available information to correctly tune \textit{k.}
In addition, the identification of equally-sized clusters contradicts some well-known facts about real social networks: the size of communities greatly varies, ranging from few communities gathering a large number of individuals, to many communities containing only few participants \cite{clauset2004finding,danon2005comparing,palla2005uncovering}.

A breakthrough in community detection has been the introduction of a cost function called {\em network modularity} (or, in short, {\em modularity}, usually denoted as \textit{Q)}; it is based on the edge density in a graph for any candidate partition. 
In the latest years, several algorithms that try to maximize \textit{Q} have been designed, see, e.g., \cite{blondel2008fast,clauset2004finding,duch2005community,newman2004fast}.

Empirical studies carried out by Newman and Girvan \cite{newman2004fast,newman2004finding} on a wide variety of artificial and real networks highlighted a correlation between high values of modularity and the actual community structure of a network.
As a further empirical result, Newman and Girvan proved that real-world networks endowed with a clear community structure have their $Q$ measure ranging from $0.3$ to $0.7.$
That means that the task of maximizing $Q$ is instrumental to disclose the community structure of a network and this is perhaps the main reason for its widespread adoption and the success of approaches that maximize modularity.

Unfortunately, the maximization of $Q$ poses two main challenges.
First of all, the $Q$ function is defined over all the vertices of the graph $G$ and, therefore, to maximize it we should take into account the whole network topology.
Approaches based on the knowledge of the whole network structure are defined as {\em global} approaches.
Among them, we cite the Girvan-Newman algorithm \cite{girvan2002community} (which is to the best of our knowledge the first algorithm that attempt to maximize \textit{Q),} the method based on information centrality \cite{FoLaMa04} and several others (please see \cite{fortunato2010community} for a survey).
The worst-case time complexity of global approaches is, unfortunately, very high.
Thus, these strategies cannot be successfully applied on very large networks comprising millions of vertices and billions of edges.

In contrast, another family of approaches considers {\em local information}, like the knowledge of the neighbors of each vertex to perform graph clustering.
These are called {\em local}approaches and one of the best-known is {\em Louvain Method -- LM} \cite{blondel2008fast}. 
In general, these approaches use greedy strategies to maximize $Q$ and, therefore, they may get trapped into local optima and, ultimately, fail to discover the actual community structure of a graph.

Second, the optimization of modularity has a major drawback called {\em resolution limit} \cite{fortunato2007resolution}, i.e., the impossibility of finding communities of small size, under certain circumstances, that typically depend on the topology of the network.
Several authors have proposed ad-hoc solutions to alleviate the resolution limit problem such as providing novel definitions of modularity \cite{li2008quantitative} or adding weights to the edges \cite{berry2011tolerating}.


In this article we both define a methodology for clustering graphs which couples the accuracy of global approaches with the computational performances guaranteed by local methods and, at the same time, mitigates the resolution limit effect.
The proposed approach is called {\em COmplex Network CLUster DEtection} or, shortly, {\em CONCLUDE}.

CONCLUDE works in two stages: in the first one, it maps graph vertices onto points of an Euclidean space and it computes the pairwise distance among them by exploiting a Euclidean-like distance metric. 
In the second stage, it adopts distances between points as a guide to perform clustering.

In the first phase, in order to map graph vertices onto points, CONCLUDE relies on the concept of  $\kappa$\textit{-path centrality} \cite{infsci2012,kbs2011}.
The $\kappa$-path edge centrality of an edge is defined as the probability that the edge is selected to spread information in the network; such probability is computed by a suitable process of information diffusion simulated by the algorithm over the network itself. 
The concept of $\kappa$-path edge centrality fits quite well with our intuitive, experience-based notion of community: groups of individuals who frequently exchange information each other while the possibility that an information goes out from the community should be regarded as an unlikely event.
Finding edges with high centrality is equivalent to disclosing preferential channels along which information flows and, ultimately, this is useful to reveal communities of individuals.

The definition of $\kappa$-path centrality is based on an information propagation model in which we assume that a message (representing a basic piece of information) is injected in an arbitrary vertex of the network and it flows along random, non-backtracking paths of length up to $\kappa$.
This model is justified by the fact that users are treated in an equally fashion and, therefore, each user is allowed to generate and spread a message.
In addition, each user has only a partial knowledge of the network  topology and, in particular, she is only aware of her neighborhood.
For these reasons, the propagation path followed by a message is {\em unpredictable} because each user can decide on her own the person to which a message has to be forwarded and, therefore, such a path can be intended as randomized.

The other assumption is that paths are non-backtracking.
In fact, we understand that a user wants to maximize the number of persons getting the message and, therefore, she has no interest in contacting her sources.
Finally, since distant users are not likely to influence each other, propagation paths are also required to be of bounded length.

The computation of $\kappa$-path edge centrality of an arbitrary edge is computationally demanding since it should consider all paths containing the given edge and, in principle, there could be exponentially-many such paths (in the number of vertices of the graph).
We propose here a heuristic algorithm, called {\em ERW-Kpath} (Edge Random Walk $\kappa$-path Centrality), to efficiently compute edge-centrality values.

Once $\kappa$-path centrality values have been computed, CONCLUDE proceeds to compute the distance between each pair of connected vertices.
Such a definition is based on the principle of {\em structural equivalence} \cite{wasserman1994social}: two vertices $i$ and $j$ are considered \textit{close} if their neighbors are close too.
In particular, a vertex $k$ which is a neighbor of both $i$ and $j$ is assumed to be close to both $i$ and $j$ if the probability that a message flows from $k$ to $i$ is comparable to the probability that it flows from $k$ to $j$.
If these probabilities are large enough, then a message received by $k$ will be forwarded, with high probability, to both $i$ and $j.$
Vice versa, if these probabilities are small enough, then a message received by $k$ is not likely to be forwarded to $i$ and $j$.
The probability that a message flows from the vertex $k$ to the vertex $i$ (resp., $j$) coincides with the centrality of the edge linking $k$ to $i$ (resp. to $j$.)

After the mapping has been performed and weights are in place, in principle we can follow it up with any off-the-shelf local algorithm for detecting communities in graphs.
In this article we discuss clustering by the Louvain Method described above, as we found it particularly suitable to be embedded into our CONCLUDE algorithm.


According to the discussion above, we can notice that CONCLUDE advances the state of the art in a number of directions:

\begin{itemize}

\item In order to compute the edge centrality values, we use non-backtracking random walks of finite length $\kappa$. 
The value of $\kappa$ can be fixed in an arbitrary fashion and, for large values of $\kappa$ (e.g., for $\kappa$ in the same order of magnitude of the graph diameter) our walks allow to explore portion of the graph which are quite far from each other. 
In this respect, the $\kappa$-path edge centrality has to be intended as an information at the global level because the horizon of the walker encompasses the whole graph. 
Interestingly, the simulation of such random walks require to select at random the starting vertex and, for each selected vertex, the \emph{walker} is asked to know the neighbors of that vertex. 
In this way, the walker is not required to know in advance the whole graph topology nor to store it.

\item Our analysis shows that the worst-case time complexity of CONCLUDE is near linear in the number of edges in the graph. 
This makes our approach computationally competitive and its performance is comparable with those of local algorithms.

\item According to our definition of distance, given three (or possibly more) vertices $i$, $j$ and $k$, these vertices are mapped onto close points if, when one of the three receives a message, then such a message is conveyed with high probability to the other two vertices.  
In this way, a group of vertices that forms a community is mapped onto a dense region of an Euclidean space. 
This ultimately makes the process of identifying communities more effective.

\item In the computation of distances, we used edge centrality as weights on edges. 
As outlined above, the procedure of weighting edges is, in general, beneficial to reduce the resolution limit problem. 
In a previous work \cite{infsci2012}, we showed that the usage of edge centrality in conjunction with some clustering algorithms raises the accuracy of these algorithms in comparison of that computed without exploiting weights on edges. 
This article, however, differs from our previous work because in \cite{infsci2012} we used a weighting procedure on edges to improve the accuracy of the clustering process. 
In this article, instead, we use edge centrality to compute distances between pairs of vertices and, as a final outcome, to map graph vertices onto points of a Euclidean space. 
Such a mapping is interesting {\em per se}  because a wide range of complex analysis tasks are allowed (like finding the $k$-nearest neighbors of a vertex).
\end{itemize}

Performance and scalability of CONCLUDE have been assessed through experiments and benchmarking both real and artificial network datasets.
In particular, we considered 6 real-world datasets where the largest, a sample from Facebook, consists of 63,731 vertices and 1,545,684 edges. 
We compared the modularity achieved by CONCLUDE clustering against the results of well-known algorithms such as the Louvain Method alone, COPRA \cite{gregory2007algorithm} and OSLOM \cite{lancichinetti2011finding}.
As for synthetic (artificially-generated) networks, we used the LFR benchmark \cite{lancichinetti2008benchmark} to generate 72 networks whose community structure was known in advance.
We compared communities found by CONCLUDE with the actual ones by using the so-called {\em Normalized Mutual Information} measure from Information Theory \cite{danon2005comparing}.
Finally, to make the algorithm available for the research community, an implementation of CONCLUDE can be now freely obtained%
\footnote{See the following URL: {\tt www.emilio.ferrara.name/CONCLUDE}}.

The outline of the article is as follows: in Section \ref{sec:background} we provide a detailed discussion of main  methods in literature to find communities in networks based on the principle of modularity maximization.
In Section \ref{sec:kpath} we provide the definition of $\kappa$-path centrality and describe our ERW-Kpath algorithm.
The main features of CONCLUDE are covered in Section \ref{sec:CONCLUDEfeatures} whereas an in-depth
experimental analysis of CONCLUDE performance is discussed in Section \ref{sec:experiments}.
Finally, in Section \ref{sec:conclusions} we draw our conclusions.

\section{Background} \label{sec:background}

In this section we first describe approaches to finding communities based on the principle of maximizing the network modularity \cite{girvan2002community} (in Section \ref{sub:networkmodularity}.)
Next, we discuss the shortcomings associated with the maximization of this function and illustrate some strategies to cope with them (Section \ref{sub:limitations}.)

\subsection{Finding Communities by Maximizing Network Modularity}\label{sub:networkmodularity}
The concept of {\em network modularity} or, in short, {\em modularity} was introduced by Girvan and Newman \cite{girvan2002community}.
Network modularity is usually denoted as \textit{Q} and it is based on the idea that a random graph is not expected to exhibit a community structure.
Therefore, for a given a graph $G = \langle V, E\rangle$ and a partition (or clustering) $\mathcal{C} = \{ \mathcal{C}_1, \ldots, \mathcal{C}_{q}\}$ of the vertices of $G$, a random graph $G^{'}$ is built by copying each vertex of $G$ onto a vertex in $G^{\prime}$.
Thus, for each vertex $i$ in $G$, there exist an homonym vertex in $G^{\prime}$.
Because of each vertex in $V$ uniquely corresponds to a vertex in $V^{\prime}$ and vice versa, for each vertex $i \in V$ we shall denote as $i^{\prime} \in V^{\prime}$ the unique vertex corresponding to $i$.
In $G^{\prime}$ an edge is drawn between a pair of vertices according to a uniform probability and, in addition, the degree sequence in $G^{\prime}$ matches the actual degree sequence of $G$.
Therefore, if a vertex $i$ has degree $d_i$ in $G$, then also $i^{\prime}$ has degree $d_i$.
Furthermore, if the total number of edges in $G$ is $m$, then the total number of edges in $G^{\prime}$ will also be $m$.
Finally, the expected number of edges linking two vertices $i^{\prime}$ and $j^{\prime}$ is proportional to the product of their degrees and, in particular, it is equal to $\frac{k_i \cdot k_j}{2m}$.
The graph $G^{\prime}$ is called the {\em null model for G.}

For a given subgraph $\mathcal{C}_i \subseteq \mathcal{C}$, we can select one by one the vertices forming $\mathcal{C}_i$ and consider, for each selected vertex, the corresponding vertex in $G^{\prime}$.
This leads to identify a subgraph $\mathcal{C}_i^{\prime}$ in $G^{\prime}$ corresponding to $\mathcal{C}_i$.
The subgraph $\mathcal{C}_i$ is classified as a community if the density of internal edges in $\mathcal{C}_i$ is significantly greater than in $\mathcal{C}_i^{\prime}$.

The $Q$ function formally encodes the previous reasoning, i.e., for each community $\mathcal{C}_i$, it checks if an edge in $\mathcal{C}_i$ exists also in $\mathcal{C}_i{^\prime}$.
This leads to the following formula

\begin{equation}
\label{eqn:qmodexp}
Q = \frac{1}{2m}\sum_{i,j} \left(A_{ij} - \frac{k_i \cdot k_j}{2m}\right) \delta(\mathcal{C}^{(i)},\mathcal{C}^{(j)}).
\end{equation}

Here $A_{ij} = 1$ if vertices $i$ and $j$ in $G$ are connected by an edge and 0 otherwise and $\mathcal{C}^{(i)}$ (resp., $\mathcal{C}^{(j)}$) is the community containing the vertex $i$.
Here $\delta(\cdot,\cdot)$ denotes the Kronecker function: two vertices $i$ and $j$ provide a non zero contribution to the value of $Q$ if and only if they belong to the same community.

The problem of maximizing $Q$ has been proved to be {\em NP-hard} by Brandes \emph{et al.} \cite{brandes2007finding}.
The first, non-trivial, approximability results beyond the NP-Hardness were proposed by Das, Gupta and Desai \cite{dasgupta2012complexity}.
They studied dense graphs separately from sparse ones and their main result proves the $\left(1 + \varepsilon\right)$-in-approximability of $Q$ in the case of dense graphs and a logarithmic approximation in the case of sparse graphs.

Several heuristic strategies to maximize the network modularity $Q$ have been proposed as to date. Probably, the most popular one is known as the \emph{Girvan-Newman strategy} \cite{girvan2002community,newman2004finding}.

In Girvan-Newman, edges are ranked by using a parameter known as {\em edge betweenness centrality}.
The edge betweenness centrality $B(e_{ij})$ of a given edge $e_{ij} \in E$ connecting the vertex $i$ with the vertex $j$ is defined as

\begin{equation}	
	B(e_{ij}) = \sum_{l \in V}\sum_{m \in V}\frac{np(l,m,e_{ij})}{np(l,m)}
	\label{eq:bc}
\end{equation}

where $l$ and $m$ are arbitrary vertices in $V$, $np(l,m)$ is the number of shortest paths connecting $l$ and $m$ and $np(l, m, e_{ij})$ is the number of the shortest paths between $l$ and $m$ containing $e_{ij}$.

It is possible to maximize $Q$ by {\em progressively deleting} edges with the highest value of betweenness centrality, based on the consideration that they shall connect vertices belonging to different communities \cite{newman2004finding}.
The process iterates until a significant increase of $Q$ is obtained.
At each iteration, each connected component of $G$ identifies a community.
Unfortunately, the computational cost of this strategy is $O(|V|^3)$ and this makes it unsuitable for the analysis of large graphs.
The most time-expensive part of the Girvan-Newman strategy is the calculation of the betweenness centrality.
Efficient algorithms have been designed to approximate the edge betweenness \cite{brandes2001faster}, or to efficiently compute shortest paths, for example in the context of weighted graphs \cite{raphael2012approximate}.
For real-world graphs, however, the computational costs still remains prohibitive.

Several variants of this strategy have been proposed during the years, such as the \textit{Fast Clustering Algorithm} provided by Clauset, Newman and Moore \cite{clauset2004finding}, that runs in $O(|V| \log |V|)$ on sparse graphs.
In \cite{duch2005community}, Duch and Arenas proposed the {\em extremal optimization method} based on a fast agglomerative approach whose worst-case time complexity is $O(|V|^2 \log |V|)$.

An interesting network modularity maximization strategy is provided in the so-called \emph{Louvain method} (LM) \cite{blondel2008fast}, which will be extensively described in Section \ref{sub:cluster-detection}.

The approaches mentioned above use {\em greedy strategies} to maximize $Q$.
In \cite{GuAm05} the authors propose to use {\em simulated annealing} to maximize $Q$.
This approach achieves a high accuracy but can as well be computationally very expensive.
The main advantage of \emph{simulated annealing} is that it does not exploit any greedy strategy, thus it is less likely to incur in the problem of becoming stuck in local optima.
Starting from a randomized partition and an arbitrary parameter that simulates the temperature of the system, the simulated annealing algorithm computes the \emph{energy} of the current configuration (say $E_a$), according to a given suitable function.
After some changes in the partitioning are applied, the algorithm recomputes the energy level of the new configuration (say $E_b$) and accepts the current solution if the new energy value is lower than the former one.
Interestingly, also a non-optimal solution might be accepted, with a probability $\Pr(\gamma)$ proportional to the \emph{temperature} $T$ of the system, computed according to the \emph{Boltzmann factor} $\Pr(\gamma) = e^{-\Delta E_{ab}/T}$ where $\Delta E_{ab} = E_b - E_a$.
Then, the algorithm lowers the temperature and iterates the process.
The lower the temperature, the lower the chance of accepting non-optimal solutions.
This means that in the early stage the algorithm tends to accept non-optimal solutions more frequently, and in the late stage the optimal solution should emerge.
This helps the algorithm to avoid getting stuck in local-optima, but contributes to its high computational cost.

\subsection{Limitations of Network Modularity}\label{sub:limitations}
In this section we discuss some limitations of the approaches attempting at optimizing $Q$.
A first problem of modularity depends on the fact that there exists an exponential number of partitions of a graph whose modularity values are close each other and these values are also close to the the global maximum of $Q$ \cite{good2010performance}.
On the one hand, such a result explains why methods which are in principle very different each other generate graph clusterings whose modularity values are quite close. On the other hand, different algorithms could produce clusterings which significantly differ each other but each of those could be associated with similar modularity scores.
The main consequence is that none of these clusterings appears to be better than others, unless we can manage additional information explaining how communities are organized.

The second (and perhaps most important) problem is known as {\em resolution limit} \cite{fortunato2007resolution}.
Informally, the resolution limit problem consists in the fact that if we hypothesize that communities of small size exist in the original graph, the value of $Q$ increases if we merge these communities into a larger one.
Merging these communities might be potentially wrong because, in order to pursue the maximization of $Q$, we would ignore small communities possibly relevant in the community structure of a graph.
From a quantitative standpoint, the resolution limit problem arises whenever in a graph $G$ there exists a community $C$ such that the sum of the degrees of vertices inside $C$ is less than $\sqrt{m}$ being $m$ the number of edges in $G$ \cite{fortunato2007resolution}.

The resolution limit problem directly depends on the definition of modularity (as reported in Equation \eqref{eqn:qmodexp}) and, in particular, from the null model exploited for defining $Q$.
In the definition of the null model we assume that an arbitrary pair of vertices in the graph can belong to the same community, independently of the position of each vertex in the graph or, in other words, that the horizon of an arbitrary vertex coincides with the whole graph.
Such an assumption can hold true for small-sized graph but it is certainly false for large scale networks like the Facebook social graph.
A possible strategy to address the resolution limit would therefore require to assume that each vertex has a {\em partial horizon}, and, consequently, it is allowed to interact with just a portion of the graph.
This would lead to consider {\em local measures} of modularity \cite{muff2005local}.

Li \emph{et al.} \cite{li2008quantitative} introduced a function, called {\em modularity density} which is not based on a null model.
The usage of modularity density is proven to yield better results than those achieved by the modularity defined in Equation \ref{eqn:qmodexp} despite it continues suffering from the resolution limit problem.
To avoid the resolution limit, the authors suggested a more general definition of modularity density which depends on a parameter $\lambda$.
The tuning of $\lambda$ allows for exploring the graph at various levels and, depending on the specific value of $\lambda$, it can favor the discovery of small communities or of large communities respectively.
An analogous study is presented in \cite{arenas2008analysis}.


Berry \emph{et al.} \cite{berry2011tolerating} studied how to alleviate the resolution limit problem in the context of weighted graphs.
In that paper, the authors suggested to assign a weight equal to 1 to each inter-cluster edge and a weight equal to $\varepsilon$ to each intra-cluster edge, being $0 < \varepsilon < 1$.
Because of this weighting procedure, the authors proved that a cluster is not detected if the following condition occurs: $w_s \leq \sqrt{\frac{W\varepsilon}{2}} - \varepsilon$, being $w_s$ the sum of the weights of intra-cluster edges, $W$ the sum of all weights of the edges in $G$ and $\varepsilon$ the maximum weight of an inter-cluster edge.
Such a result is better than the resolution limit found in \cite{fortunato2007resolution} for unweighted graphs and this means that by suitably weighting a graph it is possible to discover small-sized clusters still by optimizing modularity.

\section{The $\kappa$-path edge centrality}\label{sec:kpath}

In this section we briefly describe our method to rank edges in a graph, called $\kappa$-path edge centrality.
The $\kappa$-path edge centrality has been introduced in \cite{kbs2011} to better understand the strength of the tie bonding two vertices and, as a further step, to produce better graph clusterings \cite{infsci2012}.

The notion of $\kappa$-path centrality relies on the idea of ranking edges in a graph according to their capability of spreading information. In our approach, the basic piece of information is called {\em message}.
The introduction of $\kappa$-path edge centrality is justified by our intuitive definition of community: in fact, we assume that communities in social networks can be thought as circles of friends who frequently exchange messages each other.
Ranking edges on the basis of their aptitude to spread messages is therefore instrumental to highlight preferential pathways along which information flows and, ultimately, this is relevant to disclose communities.

To define the $\kappa$-path edge centrality we need a {\em mathematical model} describing how information flows in a network. To the best of our knowledge, one of the first models to rank edges in a graph on the basis of how information spreads and, subsequently, to discover communities was introduced by Fortunato \emph{et al.} in \cite{FoLaMa04}.
Such a model assumes that information from a source vertex to a target one is forced to travel along the shortest path connecting the two vertices. Therefore, given a pair of vertices $i$ and $j$, Fortunato \emph{et al.} defined a parameter, called {\em efficiency} as the inverse of the length of the shortest path connecting $i$ and $j$.

The hypotheses governing the model of Fortunato \emph{et al.} could not hold true in real scenarios: for instance, in online social networks like Facebook, a user is agnostic about the whole network topology and, therefore, she is not able to find shortest paths.
In addition, the computation of efficiency requires to calculate all pairs of shortest paths and this task can be prohibitively time-expensive in large networks.

To solve these drawbacks, we drop the assumption that information is forced to flow along shortest paths and assume that all paths in the graph can be exploited to convey information.
As a consequence, since all paths are eligible to convey messages, we assume that a sender selects {\em at random} one of these paths to transmit the original message.

We pose two further requirements:

\begin{enumerate}

\item {\em Simple Paths;}
We must avoid that in the information propagation process an edge is selected more than once.
This encodes the fact that, in reality, whenever a user wants to disseminate a message, she has to maximize its coverage (i.e., she wants that the message is delivered to as many users as possible.)
Therefore, she must avoid that the message is sent twice to the same user.

\item {\em Bounded Length Paths.}
As shown in \cite{friedkin1983horizons}, distant vertices in social networks (i.e., those vertices that are connected by long paths only) are unlikely to influence each other.
We agree with this observation and figure that two vertices are classified as distant if the path connecting them is longer than $\kappa$ hops, being $\kappa$ a fixed threshold.
\end{enumerate}

A path satisfying the requirements above is said {\em simple} $\kappa$-{\em path}.
We are now in the position of formally introducing the concept of $\kappa$-path centrality:

\begin{definition} ($\kappa$-path edge centrality.) {\em Let: {\em (i)} $G = \langle V, E \rangle$ be a graph, {\em (ii)}  $\kappa >0$ be an integer and {\em (iii)} $e_{ij} \in E$ be an edge of $E$ connecting the vertices $i$ and $j$.
The $\kappa$-path edge centrality $L^\kappa(e_{ij})$ of $e_{ij}$ is the sum, over all possible source vertex $s$, of the probability with which a message originated from $s$ traverses $e_{ij}$, assuming that the message traversals are only along random simple $\kappa$-paths.}\label{def:kpathedgecentrality}
\end{definition}

If we define as $\Pr(e,s)$ the probability of selecting the edge $e$ in a random simple $\kappa$-path originating from an arbitrary source vertex $s$, the centrality of an edge $e_{ij}$ reads as follows

$$
L^{\kappa}(e_{ij}) = \sum_{s \in V} \Pr(e,s).
$$

Unfortunately, Definition \ref{def:kpathedgecentrality} is hard to apply in practice because we need to consider all potential random simple paths of length at most $\kappa$ in $G$ and such a number could be exponential against the number of vertices in $G$.
To overcome this limitation, we decided to use multiple random walks to simulate the propagation of a message.
Such an idea has been already successfully exploited to compute the centrality of vertices in graphs \cite{alahakoon2011kpath,Newman2005}.

The usage of random walks to simulate simple $\kappa$-paths allowed us to design a {\em heuristic} algorithm to efficiently approximate edge centrality.
Our algorithm is called {\em ERW-Kpath - Edge Random Walk $\kappa$-path Centrality}. 
It takes as input a graph $G = \langle V, E \rangle$, an integer $\kappa$ and an integer $\rho$. 
The algorithm performs $\rho$ iterations.
We studied the impact of $\rho$ on algorithm performances in Theorem \ref{th:bounds}.

At each iteration, the ERW-Kpath algorithm works as follows:

\begin{enumerate}
\item A vertex $i \in G$ is selected as starting vertex uniformly at random among all vertices in $V$.
A message is injected in the graph starting from $i$.
All the edges in $E$ are marked as {\tt unvisited}.
A weight equal to 1 is assigned to each edge.

\item The message is propagated in the network as follows: starting from $i$, a vertex $j$ adjacent to $i$ is selected at random with uniform probability, assuming that the edge $e_{ij}$ linking $i$ and $j$ is marked as {\tt unvisited}.
If there is no edge $e_{ij}$ marked as {\tt unvisited} then the Step 2 ends.
The weight of $e_{ij}$ is increased by 1, $e_{ij}$ is marked as {\tt visited} and the propagation process restarts from $j$.
The process ends if one of the two following criteria is met: {\em (i)} a path of length $\kappa$ has been generated or {\em (ii)} there is no edge associated with $j$ which has been marked as {\tt unvisited}.

\end{enumerate}

The label associated with each edge prevents from visiting an edge more than once.
When the ERW-Kpath algorithm ends, the weight of each edge $e_{ij}$ is divided by $\rho$ and the obtained value is the edge centrality $\hat{L}^\kappa(e_{ij})$ of $e_{ij}$ returned by the ERW-Kpath algorithm.

%
%
%

The worst-case time complexity of the ERW-Kpath algorithm is $O(\kappa \rho)$ because the external loop is iterated exactly $\rho$ times whereas the internal one is carried out {\em at most} $\kappa$ times.
It is possible to show that the ERW-Kpath algorithm provides an accurate approximation of the actual centrality value of an edge.
To make the presentation of our results simpler, we shall use a simplified notation in which an edge connecting two arbitrary vertices $i$ and $j$ will be denoted as $e$ rather than $e_{ij}$ (as we did before.)

Before illustrating Theorem \ref{th:bounds}, we are in charge of defining the {\em deviation error} $\epsilon_e$ associated with an edge $e \in E$.
The deviation error is defined as follows

$$
\epsilon_ e  = \frac{|{L}^\kappa(e) - \hat{L}^\kappa(e)|}{\hat{L}^\kappa(e)}
$$

and it assess the (percentage) deviation of $\hat{L}^\kappa(e)$ to ${L}^\kappa(e)$.
The smaller $\epsilon_e$, the better the approximation of $L^\kappa(e)$.

Our goal is to show that, for an arbitrary edge $e$ and an arbitrary constant $\hat{\epsilon}$, the probability that the deviation error $\epsilon_e$ is larger than $\hat{\epsilon}$ is bounded by a function of the form $\exp^{-\rho}$.
Such a result states that a little increase in $\rho$ (and, therefore, in the running time of the ERW-Kpath algorithm) is able to produce a relevant decrease in the difference between $\epsilon_e$ and $\hat{\epsilon}$ and this ultimately implies that the algorithm is able to quickly converge to the actual values of edge centrality.

To prove such a property we need a preliminary result known as {\em Hoeffding inequality}:

\begin{theorem}{(Hoeffding inequality)}
{\em Let $X_1, \ldots, X_n$ be independent random variables.
Assume that, for each $i$ such that $1\leq i \leq n$, the random variable $X_i$ ranges in the real interval $\left[ a_i,b_i \right]$.
Let $\overline{X} = (X_1 + \cdots + X_n)/n$.
For any $t \geq 0$ we have
\begin{equation}
	\label{eqn:hoeffdinggeneral}
	\Pr(|\overline{X} - \mathrm{E}[\overline{X}]| \geq t) \leq 2\exp \left( -
	\frac{2t^2n^2}{\sum_{i=1}^n (b_i - a_i)^2} \right).
\end{equation}}
\end{theorem}

{\em Proof}. See \cite{Hoeffding63} \begin{flushright} $\Box$ \end{flushright}

As a special case, if all random variables $X_i$ can only take up value 0 or 1 then Equation \eqref{eqn:hoeffdinggeneral} simplifies to

\begin{equation}
\label{eqn:hoeffdingsimpl}
\Pr(|\overline{X} - \mathrm{E}[\overline{X}]| \geq t) \leq 2\exp \left( -
2t^2n\right).
\end{equation}

We are now able to prove our claims.

\begin{theorem}
\label{th:bounds} {\em Let: {\em (i)} $G = \langle V, E \rangle$ be a graph, {\em (ii)} $\langle \kappa, \rho \rangle$ be a pair of positive integers and {\em (iii)} $\hat{\epsilon}$ a positive real number.
Assume to run the ERW-Kpath algorithm on $G$ so that the algorithm generates $\rho$ simple paths of length up to $\kappa$ and let $\hat{L}^{\kappa}(e)$ be the edge centrality value of an arbitrary edge $e \in E$ returned by the ERW-Kpath algorithm.
The probability that the deviation error $\epsilon_e$ is greater than or equal to $\hat{\epsilon}$ is bounded by $e^{-\rho}$.
}
\end{theorem}

{\em Proof}. By Definition \ref{def:kpathedgecentrality} and the definition of conditional probability, we get

\begin{equation}
\label{eqn:edgecentralityformula}
L^\kappa(e) = \sum_{s\in V}\Pr(e,s) = \sum_{s\in V}\Pr(e|s)\Pr(s) = \frac{1}{|V|}\sum_{s\in V}\Pr(e|s)
\end{equation}

In Equation \ref{eqn:edgecentralityformula}, the term $\Pr(s)$ is the probability that $s$ is selected as the source vertex and it equals $\frac{1}{|V|}$ because the starting vertex is selected with uniform probability across all vertices in $V$.

The ERW-Kpath algorithm performs $\rho$ iterations, and, at each iteration, it generates a simple $\kappa$-path. In our analysis we will first focus on the result produced in a given iteration, say the $\ell$-th iteration with $1 \leq \ell \leq \rho$.

Let us define the random variable $Y_{\ell}(e)$ as follows

$$
Y_{\ell}(e) = \left\{
			\begin{aligned}
				& 1 \quad  \mbox{if $e$ has been selected at the $\ell$-th iteration}\\
                & 0 \quad \mbox{otherwise.}
			\end{aligned}\right.
$$

Since $Y_{\ell}(e)$ is an indicator variable, we have that $E[Y_{i}(e)] = \Pr(Y_{i}(e) = 1)$.

For a fixed $\ell$, $\Pr(Y_\ell(e)= 1)$ is equal to the probability of selecting $e$ independently of the choice of starting vertex $s$.
Therefore, if we denote as we did before $\Pr(e,s)$ the probability that ERW-Kpath algorithm selects $e$ when the starting vertex is $s$ we get

$$
\Pr(Y_i(e)= 1) = \sum_{s \in V} \Pr(e,s) = \sum_{s \in V} \Pr(e|s)\Pr(s) = \frac{1}{|V|}\sum_{s \in V} \Pr(e|s)
$$

The ERW-Kpath algorithm increases by 1 the weight of $e$ each time $e$ is selected and, at the end, it divides the total weight by $\rho$.
It follows that the value of centrality $\hat{L}^\kappa(e)$ returned by the ERW-Kpath algorithm is a random variable distributed as $\frac{1}{\rho} \sum_{\ell = 1}^{\rho} Y_\ell(e)$.
The average of  $\hat{L}^\kappa(e)$ is as follows

$$
E\left[\hat{L}^\kappa(e)\right] = \frac{1}{\rho} \sum_{\ell = 1}^{\rho} E[Y_\ell(e)] =
\frac{1}{\rho} \sum_{\ell = 1}^{\rho} \Pr[Y_\ell(e) = 1] = \frac{1}{\rho} \sum_{\ell = 1}^{\rho}\left( \frac{1}{|V|}\sum_{s \in V} \Pr(e|s)\right) = \frac{1}{\rho}\times \rho \times {L}^\kappa(e) = {L}^\kappa(e)
$$

From Equation \ref{eqn:hoeffdingsimpl}, with $t = \hat{\epsilon} \hat{L}^\kappa(e)$ it follows that

$$
\Pr\left(\left| L^\kappa(e) - \hat{L}^\kappa(e) \right| \geq \hat{\epsilon} \hat{L}^\kappa(e) \right) \leq 2\exp\left(-2\rho\hat{\epsilon}^2 \hat{L}^\kappa(e)^2\right)
$$

Since $\hat{L}^\kappa(e) \geq \frac{1}{\rho} > 0$, the previous equation can be rewritten as follows

$$
\Pr\left(\frac{\left| \hat{L}^\kappa(e) - L^\kappa(e) \right|}{L^\kappa(e)} \geq \hat{\epsilon} \right) \leq 2\exp\left(-2\rho\hat{\epsilon}^2 \hat{L}^\kappa(e)^2\right) \Rightarrow \Pr\left(\epsilon_e \geq \hat{\epsilon} \right) \leq 2\exp\left(-2\rho\hat{\epsilon}^2 \hat{L}^\kappa(e)^2\right) =  K \exp(-\rho)
$$

where $K = \exp\left(2 \hat{\epsilon}^2 \hat{L}^\kappa(e)^2\right)$.
The latest equation states that the probability that the deviation error associated with an arbitrary edge $e$ is greater than $\hat{\epsilon}$ is bounded (up to a constant factor $K$) by the function $e^{-\rho}$ and this ends the proof. \begin{flushright} $\Box$ \end{flushright}

According to Theorem \ref{th:bounds}, we can conveniently fix $\rho$ so that to make $\Pr\left (|\hat{L}^{\kappa}(e) - L^{\kappa}(e)| \geq \epsilon \right)$ for an arbitrary threshold $\epsilon$ as little as we need.
In particular, for a fixed $\epsilon$, we can set $\rho = \frac{\alpha \log |V|}{2\epsilon^2\hat{L}^{\kappa}(e)^2}$, being $\alpha$ any real number greater than 1 to obtain the following result

$$
\Pr\left(\frac{\left| L^\kappa(e) - \hat{L}^\kappa(e) \right|}{\hat{L}^\kappa(e)} \geq \epsilon \right) \leq\exp\left(-2\rho\hat{\epsilon}^2 \hat{L}^\kappa(e)^2\right) = \exp\left(-2 \frac{\alpha \log |V|}{2\epsilon^2\hat{L}^{\kappa}(e)^2}\epsilon^2 \hat{L}^\kappa(e)^2\right) =
\exp(-\alpha \log |V|) = \frac{1}{|V|^\alpha}
$$

In such a case, the worst-case time complexity of the ERW-Kpath algorithm is $O(\kappa \alpha \log |V|)$ and its deviation error (for each edge $e \in E$) is no larger than $\frac{1}{|V|^\alpha}$.

\section{Computing Distances and Finding Communities}\label{sec:CONCLUDEfeatures}
In this section we describe how to use $\kappa$-path edge centrality to compute distances between vertices in a graph (see Section \ref{sub:proximity}); after this, we show how to use these distances in conjunction with the Louvain Method to identify communities in a graph (see Section \ref{sub:cluster-detection}.)
Finally, we compare our method to a measure of centrality that turns out to be very close to ours (see Section \ref{sub:comparison}.)

\subsection{Computing Distances between Graph Vertices}\label{sub:proximity}
Once $\kappa$-path centrality values have been computed, CONCLUDE proceeds to calculate the distances between each pair of vertices; computing distances allows for embedding graph vertices into a metric space.
Such a feature is {\em per se} interesting, independently of the fact we used such an embedding to perform graph clustering.

In fact, in many Database and Information Retrieval Applications, a large (and perhaps one of the most common) class of queries assumes that a collection of data items is available and requires to find the items in the collection best matching with a particular query item, according to some definition of best.
For example, given a database reporting images, a typical query is to find all images that are similar to a particular query image.
If a metric distance can be defined between pair of items composing the database, then it is possible to design efficient distance-based data
structures which make the retrieval or the indexing of that items easy and fast even on huge collections \cite{chavez2001searching}.

Our approach to compute distances relies on the computation $\kappa$-path edge centralities and this provides two main advantages.
First, the computation of distances is fully automatic and does not require any human intervention.
Second, our definition of distance is based on the information propagation model discussed in Section \ref{sec:kpath}.
This provides our definition of distance with a meaningful interpretation: two points are as much distant as the chance they can exchange messages is low.
In particular, two vertices $i$ and $j$ are deemed close if a message received by $i$ is sent, with high probability, to the vertex $j$ and vice versa.
To formally encode such a reasoning, we could first define a {\em proximity measure} $\sigma_{ij}$ between the vertices $i$ and $j$ as the $\kappa$-path centrality of the edge linking $i$ and $j$.
Unfortunately, such a measure of proximity is not satisfactory for two reasons: first of all, no edge could exist between vertices $i$ and $j$ and, in such a case, the proximity $\sigma_{ij}$ would be undefined.
Even assuming the existence of an edge from $i$ to $j$ and an edge from $j$ to $i$, it can happen that the centrality of the edge going from $i$ to $j$ differs from the centrality of the edge linking $j$ and $i$ and this implies that $\sigma_{ij} \neq \sigma_{ji}$, thus violating the symmetry requirement imposed by the concept of proximity.

To overcome these limitations, we rely on the principle of {\em structural similarity} \cite{wasserman1994social} and we guess that two vertices $i$ and $j$ are to be considered as close if their neighbors are close too.
In line with the previous reasoning, a vertex $k$ is close to both $i$ and $j$ if the probability that a message flows from $k$ to $i$ is comparable to the probability that it flows from $k$ to $j$.
The same definition implies that the vertices $i$ and $j$ are classified as distant if the probability that a message flows from $k$ to $i$ is high (resp., low) and the probability that a message flows from $k$ to $j$ is low (resp., high.)
The probability of conveying a message from $k$ to $i$ (resp, $j$) coincides with the $\kappa$-path edge centrality of the edge $e_{ki}$ (resp., $e_{kj}$.)
Such a definition is quite interesting because the vertices $i$, $j$ and $k$ would form a triangle in which, if one of the three vertices receives a message, then such a message is conveyed with high probability to the two other vertices; in an analogous fashion, if no one of these vertices receives a message, the chance that the two other vertices will receive it is low.

By generalizing the previous concept, we can consider a group $\mathcal{C}$ of vertices in $G$ and we can observe that if a message flows with high probability among the vertices in $\mathcal{C}$ then $\mathcal{C}$ is mapped onto a dense region and this highlights that $\mathcal{C}$ is a community in $G$.
This argument leads us to consider the following definition of proximity

\begin{equation}
\label{eq:sigma}
\sigma_{ij} = \sqrt{\sum_{k \in V}\left[L^{\kappa}(e_{ki}) - L^{\kappa}(e_{kj})\right]^2}
\end{equation}

Equation \ref{eq:sigma} shows some nice properties: it uses the contribution of all vertices in $V$ to quantify the proximity degree of vertices $i$ and $j$; in addition, it is based on the traditional Euclidean distance definition and therefore, it maps vertices of $G$ onto points of an Euclidean space.

In Equation \ref{eq:sigma}, we assume that all vertices are equally important in deciding whether two vertices are close, regardless of their degrees. Such a choice, however, can yield unwanted results because the degree of vertex can influence the centrality of edges incident onto it.

To convince ourselves of this fact, let us select at random a vertex $s \in V$ and assume to generate a random simple $\kappa$-path $p_s = \{ s, i_1, i_2, \ldots, i_{\kappa -1} \}$ starting from $s$.
Let us focus on a single run of the ERW-Kpath algorithm and observe that the probability of selecting the edge connecting $s$ and $i_1$ is equal to the ratio $\frac{1}{|N(s)|}$, being $N(s)$ the set of neighbors of the vertex $s$.
Analogously, the probability of going from $i_1$ to $i_2$ is equal to $\frac{1}{|N(i_1) - \{s\}|}$.
The previous formula is justified by the fact that the path generated in a single run by the ERW-Kpath algorithm cannot pass twice for the same edge: therefore if $i_2$ would coincide with $s$, the edge connecting $i_2$ with $i_1$ cannot be selected. In such a case the vertex $i_2$ must be excluded from $N(i_1)$.


By generalizing the previous result, the if we consider a path $p_s = \{s, i_1, \ldots, i_l \}$ and we focus on two vertices $i_{a}$ and $i_{a+1}$, the probability to select the edge joining $i_{a}$ and $i_{a+1}$ is

$$
\Pr(e_{i_a i_{a+1}}) = \left\{
			\begin{aligned}
				& 0 \quad \mbox{if $N(i_{a}) \subseteq \{s, i_1, \dots, i_{a-1}\}$} \\
                & \frac{1}{|N(i_{a}) - \{s, i_1, \dots, i_{a-1}\}|} \quad  \mbox{otherwise.}
            \end{aligned}
            \right.
$$

Intuitively, the special case $\Pr(e_{i_a i_{a+1}}) = 0$ occurs when all edges incident on $N(i_{a})$ have been already visited during the considered path $\{s, i_1, \dots, i_{a-1}\}$.
We denote with $deg(i_a)$ the degree of $i_a$ and observe that $\mathrm{deg}(i_a) = |N(i_a)| \geq |N(i_{a}) - \{s, i_1, \dots, i_{a-1}\}| \geq |N(i_{a}) - \kappa|$. In practical scenarios, the value of $\kappa$ is intended as  small and, therefore, $\Pr(e_{i_a i_{a+1}})$ is proportional to $\frac{1}{\mathrm{deg}(i_a)}$.

According to these observations, the probability of selecting an edge is biased by the degree of the vertex associated with it. Therefore, the larger the degree of a vertex $i$, the easier is that a message passes through $i$ in several simulations. Therefore, $i$ would be classified as close to almost all vertices.

%
%
%

This suggests us the idea of alleviating the dependency of the proximity calculation on the degrees of the considered vertices. To do so, we suggest to normalize the proximity as follows

\begin{equation}
	\sigma_{ij} = \sqrt{\sum_{k \in V}{\frac{[L^{\kappa}(e_{ki}) - L^{\kappa}(e_{kj})]^2}{\mathrm{deg}(k)}}}
	\label{eq:distance}
\end{equation}

\noindent and, subsequently, to define a the distance measure between vertices as $d_{ij} = 1 - \sigma_{ij}$.


The naive implementation of the distance computation could be time-expensive as it requires $O(|V|^2)$ iterations.
Yet its cost can be decreased to nearly linear by observing that we can rewrite Equation \ref{eq:distance} as

\begin{equation}
	\sigma_{ij} = \sqrt{ \frac{ \displaystyle \sum_{k \in N(i) - CN(i,j)}{[L^\kappa (e_{ki})]^2 } }{ | N(i) - CN(i,j) | } +
	\frac{ \displaystyle \sum_{k \in N(j) - CN(i,j)}{[L^\kappa (e_{kj})]^2} }{ | N(j) - CN(i,j) | } +
	\frac{ \displaystyle \sum_{k \in CN(i,j)}{[L^\kappa (e_{ki}) - L^\kappa (e_{kj})]^2} }{ | CN(i,j) | }}
	\label{eq:eff-distance}
\end{equation}
where the symbol $CN(i,j)$ indicates the subset of neighbors common to $i$ and $j$.

By adopting this shrewdness, the cost of the computation is reduced to $O(\overline{d}(v)^2|V|)$, where $\overline{d}(v)$ is the average degree of the vertices of the network.
%
%
%
%

\subsection{Finding Communities}\label{sub:cluster-detection}
The last step of our method consists in the network partition.
CONCLUDE adopts the paradigm of the \emph{network modularity} maximization which we described in detail in Section \ref{sec:background} and exploits an approximate technique inspired by the \emph{Louvain method} (LM) \cite{blondel2008fast}.

LM is an computationally effective algorithm, thus it is well-suited for partitioning large networks.
It consists of two steps which are iteratively repeated. The input of LM is a weighted network $G = \langle V, E, W \rangle$ being $W$ the weights associated with each edge\footnote{Of course, in case of unweighted graphs, $W$ is the adjacency matrix of $G$.}. The modularity is defined as in Equation \eqref{eqn:qmodexp}, in which $A_{ij}$ is the weight of the edge linking $i$ and $j$ and $k_i$ (resp., $k_j$) is the sum of the edges incident onto $i$ (resp., $j$.)

Initially, each vertex $i$ forms a community and therefore, there are as many communities as vertices in $V$.
%

The two steps of the LM operate as follows:

\emph{(i)} For each vertex $i$, the algorithm computes the gain $\Delta Q$ derived from moving $i$ to a cluster $C$, as

\begin{equation}
	\Delta Q= \frac{\sum_{C} + k_i^C}{2m} - \left( \frac{\sum_{\hat{C}} + k_i}{2m} \right)^2
	-\left[ \frac{\sum_{C}}{2m} - \left( \frac{\sum_{\hat{C}}}{2m} \right)^2 - \left( \frac{k_i}{2m} \right) \right]	
	\label{centrality:eq:deltaq}
\end{equation}

Here $\sum_{C}$ is the sum of the weights of the edges inside $C$, $\sum_{\hat{C}}$ is the sum of the weights of the edges incident onto vertices in $C$, $k_i$ is the sum of the weights of the edges incident to the vertex $i$, $k_i^C$ is the sum of the weights of the edges from $i$ to vertices located inside $C$ and $m$ is the sum of the weights of all the edges in the network.

The vertex $i$ is placed in the cluster $\overline{C}$ for which the gain achieves its maximum value. If it is not possible to achieve a positive gain, the vertex $i$ will remain in its original community. This process is applied repeatedly and sequentially for all the vertices until no further improvement can be achieved.

\emph{(ii)} In the second step, the algorithm builds a meta-network whose vertices are those clusters found in Step {\em (i)}, collapsing all edges among vertices belonging to a pair of cluster onto a single edge. The weight of this edge is equal to the sum of the weights of the collapsed edges. Once the second step has been performed, the algorithm re-applies the first step.

Steps {\em (i)} and {\em (ii)} are repeated until an arbitrarily small improvement $\Delta Q$ is attained at each iteration. The cost of the whole process is $O(\gamma|V|)$, where $\gamma$ is the number of iterations required by the algorithm to converge (in our experience, $\gamma < 5$.)

The advantage of our approach with respect to the original LM is twofold: first, we obtain the splitting of clusters connected by edges with low distance, which is a global feature, maximizing the network modularity, while LM only relies on local information (i.e., vertex neighborhood); second, our strategy is able to produce an edge weighting, while the original LM -- and most of current clustering algorithms -- cannot infer edge weights in case of unweighted networks. 
This aspect ensures better performance of our strategy in most of cases (as discussed in the remainder of this article.)

Summarizing, by adopting efficient graph memoization techniques, the computational cost of CONCLUDE is near linear.
In fact, it results from the three previously described steps, i.e., $O(\kappa |E| + \overline{d}(v)^2 |V| + \gamma |V|) = O(\Gamma |E|)$.

The pseudo-code describing the various steps performed by CONCLUDE is reported in Algorithm \ref{alg:CONCLUDE}.
In detail, CONCLUDE takes as input the graph $G$, an integer $\kappa$ and an integer $\rho$.
It first calls a subroutine \textit{ERW-Kpath} that computes $\kappa$-path edge centralities on vertices of $G$ by performing at most $\rho$ iteration.
The output of this subroutine is stored in an array of weights $\vec{\omega}$.
In the second step CONCLUDE calls a subroutine \textit{Compute-Pairwise-Distances} which takes as input the graph $G$ along with the array of weights $\vec{\omega}$ and computes distances among all pairs of vertices by applying Equation \ref{eq:eff-distance}.
The subroutine \textit{Compute-Pairwise-Distances} returns as output a matrix $\Delta$ containing all pairs of distances between vertices.
Finally, in its final step CONCLUDE calls the subroutine \textit{Louvain-Method,} which implements the LM method.
It takes as input the matrix $\Delta$ and returns as output the community structure $\mathcal{C}$ of $G.$

\begin{algorithm}
	\caption{CONCLUDE($G=\  \langle V,E \rangle$: a Graph, $\kappa$: an integer, $\rho$: an
integer)}
	\label{alg:CONCLUDE}
	\begin{algorithmic}[1]
		\STATE $\vec{\omega} \leftarrow \hbox{ERW-Kpath}(G, \kappa, \rho)$
		\STATE $\Delta \leftarrow \hbox{Compute-Pairwise-Distances}(G, \vec{\omega})$
		\STATE $\mathcal{C} \leftarrow \hbox{Louvain-Method}(G, \Delta)$
	\end{algorithmic}
\end{algorithm}

\subsection{Comparison with existing definition of distances}\label{sub:comparison}
In the past, many papers have focused on the problem of computing distances between vertices of a graph and, subsequently, to use such a distance with the purpose of clustering the graph.
A nice approach is due to Donetti and Mu\~{n}oz \cite{donetti2004detecting}.
In that paper, the authors suggested to consider the {\em Laplacian} $\mathbf{L}_G$ associated with a graph $G$, which is defined as $\mathbf{L}(G) = \mathbf{D}_G  - \mathbf{A}_G$.
Here $\mathbf{D}_G$ is a diagonal matrix such that $\mathbf{D}_G[i,i]$ is equal to the degree of the $i$-th vertex and $\mathbf{A}_G$ is the adjacency matrix of $G$.
The graph $G$ is assumed undirected and, therefore, $\mathbf{A}_G$ is a symmetric matrix.
The approach of \cite{donetti2004detecting} suggests to compute the first $D$ non-trivial eigenvectors of $\mathbf{L}(G)$, i.e., the eigenvectors corresponding to the $D$ largest and non-zero eigenvalues of $\mathbf{L}(G)$.
Subsequently, graph vertices are projected onto the subspace spanned by these eigenvectors.
In this way each vertex is represented by a point in $\mathbb R^D$ and the distance between two vertices is defined as the distance of the two vectors in $\mathbb R^D$ corresponding to them.

To make the comparison between our approach and that of Donetti and Mu\~{n}oz \cite{donetti2004detecting} fair,  we need to assume that $G$ is symmetric.
Under this conditions, if we relax the hypothesis of non-backtracking paths we can give a {\em spectral interpretation} of $\kappa$-path edge centrality values computed by our ERW-Kpath algorithm.
The ERW-Kpath algorithm initially builds a $n \times n$ matrix $\mathbf{E}_G$.
The generic entry $\mathbf{E}_G[i,j]$ is equal to 0 if there is no edge going from the vertex $i$ to the vertex $j$ and it is equal to $1$ otherwise.
By construction, the matrix $\mathbf{E}_G$ is symmetric and, therefore, there is an {\em orthogonal family} of eigenvectors of $\mathbf{\vec{e}}_1, \ldots, \mathbf{\vec{e}}_n$ associated with $\mathbf{E}_G$ \cite{meyer2000matrix}.
This means that $\mathbf{\vec{e}}_i^t \cdot \mathbf{\vec{e}}_j = \delta_{ij}$, being $\delta_{ij}$ the above mentioned Kronecker function.
We can write $\mathbf{E}_G$ as $\mathbf{E}_G  = \sum_{i = 1}^{n} \lambda_i \mathbf{\vec{e}}_i \mathbf{\vec{e}}_i^t$, being $\lambda_i$ the $i$-th eigenvalue.
The probability of going from the vertex $l$ to the vertex $m$ with a path of length 2 can be obtained by computing $\mathbf{E}_G^2 = \mathbf{E}_G \mathbf{E}_G $ and by picking $\mathbf{E}_G^2[l,m]$.

With some simple calculations we get

$$
\mathbf{E}_G^2 = \left(\sum_{i = 1}^{n} \lambda_i \mathbf{\vec{e}}_i \mathbf{\vec{e}}_i^t\right)\left(\sum_{j = 1}^{n} \lambda_j \mathbf{\vec{e}}_j \mathbf{\vec{e}}_j^t\right) = \sum_{i,j= 1}^{n} \left(\lambda_i \mathbf{\vec{e}}_i \mathbf{\vec{e}}_i^t\right) \left(\lambda_j \mathbf{\vec{e}}_j \mathbf{\vec{e}}_j^t\right) = \sum_{i,j= 1}^{n} \lambda_i \lambda_j \mathbf{\vec{e}}_i \left(\mathbf{\vec{e}}_i^t \mathbf{\vec{e}}_j\right) \mathbf{\vec{e}}_j^t =  \sum_{i,j= 1}^{n} \lambda_i \lambda_j \mathbf{\vec{e}}_i \delta_{ij} \mathbf{\vec{e}}_j^t = \sum_{i= 1}^{n} \lambda_i^2 \mathbf{\vec{e}}_i \mathbf{\vec{e}}_i^t
$$

\noindent
and, more in general, $\mathbf{E}_G^p = \sum_{i= 1}^{n} \lambda_i^p \mathbf{\vec{e}}_i \mathbf{\vec{e}}_j^t$.
In these products, the larger $\lambda_i$ is, the higher its contribution to $\mathbf{E}_G$.
As in \cite{donetti2004detecting}, we could pick only the $D$ largest eigenvalues to get an approximation of $\mathbf{E}_G^p$.
The probability that ERW-Kpath selects an edge depends on matrices $\mathbf{E}_G^p$ (with $p \leq \kappa$), and therefore, the edge centrality of a vertex is influenced by the top $D$ non-trivial eigenvalues of $\mathbf{E}_G$.
According to Equation \ref{eq:eff-distance}, the distances computed by our approach depends on edge centralities, and, therefore, we obtain that also distances in our approach are influence by the $D$ eigenvalues having the largest magnitude.

From a computational standpoint, \cite{donetti2004detecting} uses the popular Lanczos algorithm to find the eigenvalues of the Laplacian which, {\em de facto}, computes the power of a matrix by applying some efficient procedures like the Gram-Schmidt algorithm. Lanczos algorithm is efficient over sparse matrices but, unfortunately, to the best of our knowledge, there is no theoretical estimation of its worst-case complexity. Our approach, instead, takes $O(\kappa \alpha \log |V|)$ steps.

\section{Experimental Results} \label{sec:experiments}
CONCLUDE has been experimented in different fields of application, for example to analyze \emph{online social networks} and \emph{biological networks}.
In this article we discuss its application to the following two cases (see Section \ref{sec:social-networks}): \emph{(i)} artificially-generated (henceforth, synthetic) networks with a pre-defined community structure; \emph{(ii)} network datasets from real-world applications.
We also experimentally studied the behavior of ERW-Kpath algorithm to rank edges according to their ability of spreading messages in a network.



\subsection{Cluster Detection}\label{sec:social-networks}
In order to evaluate the performance of our clustering method, we carried out two different types of experiments.
The former, by using synthetic networks for which a pre-built community structure was well defined.
The latter, by considering different graphs describing the structure of some real-world networks.

The purpose of the former set of experiments is to assess the quality of the clustering produced by our algorithm in the context of a controlled environment in which the features of the networks taken into account are well-known.
In particular, the pre-defined community structure is adopted as a \emph{ground truth} to evaluate the quality of the partitions obtained by using our algorithm, and this is done by adopting a measure inherited by information theory (called \emph{normalized mutual information}) to establish the accuracy of the partitions with respect to the ground truth.

The second set of experiments, instead, is carried out to evaluate the performance of our method compared against three other state-of-the-art algorithms in real-world applications.
This comparison is performed by considering network datasets whose correct community structure is not known in advance; to assess the validity of obtained results we exploit the internal quality measure of \emph{network modularity} previously presented in Section \ref{sub:networkmodularity}.
In fact, regardless the model adopted to finding communities in a network, the \emph{network modularity} has been commonly adopted to establish if a community detection method has been able to discover clusters of vertices tightly interconnected among each other and loosely interconnected with those belonging to other clusters.
We assume that, the higher the values of \emph{network modularity} provided by an algorithm, the better the community structure of the network has been unveiled.
Given the limitations of the \emph{network modularity} -- as discussed in Section \ref{sub:limitations} -- this approach is only a best-approximation of any robust evaluation method.
Indeed, the problem of evaluating the clustering quality of real-world networks lacking of a ground truth is an open and urgent problem in current literature.

\subsubsection{Synthetic networks}
The first set of experiments has been carried out by using the LFR benchmark presented by Lancichinetti \emph{et al.} \cite{lancichinetti2008benchmark}.
In the LFR benchmark a user is required to provided the following information to generate a graph $G$: {\em (i) Number of Vertices}, denoted as $N$, and {\em Average Vertex Degree}, denoted as $\langle k \rangle$. {\em (ii) Power Law exponent in vertex degree distribution} denoted as $\gamma$.
The vertex degree distribution in $G$ is shaped as $k^{-\gamma}$. {\em (iii) Power Law exponent in community size distribution} denoted as $\beta$.
The size of communities in $G$ follows a power law distribution with exponent equal to $\beta$.
The sum of the sizes of all the communities is constrained to be equal to $N$. {\em (iv) Mixing parameter}, denoted as $\mu$.
A user can specify a parameter $\mu \in (0,1)$ such that each vertex in $G$ shares a fraction $1 - \mu$ of its edges with vertices outside its community and $\mu$ edges with vertices residing in its community.
The value $\mu=0.5$ is the threshold beyond which clusters are no longer defined in the strong sense (that is that each vertex has more neighbors in its own cluster than in the others.)

To generate our benchmark networks, we adopted the same configuration reported in \cite{lancichinetti2008benchmark}:
\emph{(i)} $N=1000$; \emph{(ii)} four pairs of values for $\gamma$ and $\beta$ defined as follow: $(\gamma,\beta)$ = (2,1), (2,2), (3,1), (3,2);
\emph{(iii)} for each pair of exponents $\gamma$ and $\beta$, three values of average degree $\langle k \rangle = 15,20,25$;
\emph{(iv)} for each of the previous combinations, we generated six networks by varying the mixing parameter $\mu = 0.1, \dots, 0.6$.

To compute the quality of the results, we adopted the measure called \emph{normalized mutual information} (NMI) \cite{danon2005comparing}.
Such a measure assumes that, given a graph $G$, a {\em ground truth} is available to verify what are the clusters (said {\em real clusters}) in $G$ and what are their features.
Let us denote as $A$ the true community structure of $G$ and suppose that $G$ consists of $c_A$ clusters.
Let us consider a clustering algorithm applied on $G$ and assume that it identifies a community structure $B$ consisting of $c_B$ clusters.
We define a $c_A \times c_B$ matrix -- called {\em confusion matrix} $CM$ -- such that each row of $CM$ corresponds to a cluster in $A$ whereas each column of $CM$ is associated with a cluster in $B$.
The generic element $CM_{ij}$ is equal to the number of elements of the real $i$-th cluster which are also present in the $j$-th cluster found by the algorithm.
Starting by this definition, the \emph{normalized mutual information} is defined as

\begin{equation}
\label{eqn:mutualinformationdef}
NMI(A,B) = \frac{-2\sum_{i = 1}^{c_A}\sum_{j = 1}^{c_B} N_{ij} \log \left(\frac{N_{ij}N}{N_{i\cdot}N_{\cdot j}} \right)}{\sum_{i=1}^{c_A}N_{i \cdot} \log \left(\frac{N_{i \cdot}}{N}\right) + \sum_{j=1}^{c_B}N_{\cdot j} \log \left(\frac{N_{\cdot j}}{N}\right)}
\end{equation}

where $N_{i \cdot}$ (resp., $N_{\cdot j}$) is the sum of the elements in the $i$-th row (resp., $j$-th column) of the confusion matrix.
If the considered clustering algorithm would work perfectly, then for each discovered cluster $j$, it would exist a real cluster $i$ exactly coinciding with $j$.
In such a case, it is possible to show that $NMI(A,B)$ is exactly equal to 1 \cite{danon2005comparing}.
By contrast, if the clusters detected by the algorithm are totally independent of the real communities then it is possible to show that the NMI is equal to 0.
The NMI, therefore, ranges from 0 to 1 and the higher the value, the better the clustering algorithm performs with respect to the the ground truth.

The performance provided by CONCLUDE, reported in Table \ref{tab:synthetic}, shows excellent values of NMI considering the task of solving the LFR benchmark.
For each configuration, the partition provided by our algorithm is compared against the \emph{ground truth} built by the LFR benchmark, measuring the goodness of the partition according to the previously-defined \emph{normalized mutual information} measure.

CONCLUDE provides in general high values of NMI for the setting of $\mu = 0.1, \dots, 0.3$, which lead to the presence of strongly defined clusters in the synthetic networks.
Moreover, it is worth to note that the values of NMI are stable across different configurations.
Fixed the \emph{mixing parameter} $\mu$, let the parameters $\langle k \rangle$, $\gamma$ and $\beta$ vary; their variation reflects on the feature of the networks generated by the benchmark.
We observe, in this scenario, that our strategy works well and consistently according to different network structures and that the performance is independent of particular network features (such as the degree distribution or the size of clusters present in the network.)

\begin{table}[!ht]
	\centering
	\begin{tabular}{| c c c c c c c |}
		\hline \hline
		\ $\langle k \rangle$ \ &	\ $\mu = 0.1$	\ &	\ $\mu = 0.2$	\ &	\ $\mu = 0.3$	\ &	\ $\mu = 0.4$	\ &	\ $\mu = 0.5$	\ & \ $\mu = 0.6$ \ \\
		\hline \hline
		\multicolumn{7}{|c|}{$\gamma = 2, \beta = 1$}\\
		\hline
		$15$	 & 0.861	& 0.761	& 0.664	& 0.559	& 0.409	& 0.281\\
		$20$	 & 0.859	& 0.759	& 0.663	& 0.566	& 0.434	& 0.298\\
		$25$	 & 0.856	& 0.760	& 0.663	& 0.560	& 0.433	& 0.307\\
		\hline \hline
		\multicolumn{7}{|c|}{$\gamma = 2, \beta = 2$}\\
		\hline
		$15$	 & 0.856	& 0.757	& 0.664	& 0.550	& 0.463	& 0.317\\
		$20$	 & 0.859	& 0.763	& 0.665	& 0.555	& 0.467	& 0.316\\
		$25$	 & 0.861	& 0.761	& 0.664	& 0.569	& 0.469	& 0.340\\
		\hline \hline
		\multicolumn{7}{|c|}{$\gamma = 3, \beta = 1$ }\\
		\hline
		$15$	 & 0.851	& 0.761	& 0.636	& 0.522	& 0.348	& 0.238\\
		$20$	 & 0.863	& 0.768	& 0.662	& 0.517	& 0.409	& 0.248\\
		$25$	 & 0.860	& 0.762	& 0.665	& 0.562	& 0.422	& 0.283\\
		\hline \hline
		\multicolumn{7}{|c|}{$\gamma = 3, \beta = 2$}\\
		\hline
		$15$	 & 0.865	& 0.767	& 0.657	& 0.569	& 0.407	& 0.287\\
		$20$	 & 0.863	& 0.768	& 0.670	& 0.570	& 0.435	& 0.297\\
		$25$	 & 0.863	& 0.765	& 0.666	& 0.568	& 0.448	& 0.281\\
		\hline \hline
	\end{tabular}
	\caption{Values of \emph{normalized mutual information} provided by CONCLUDE resolving the clusters in the networks with community structure artificially-generated according to the LFR benchmark \cite{lancichinetti2008benchmark}.}
	\label{tab:synthetic}
\end{table}

\subsubsection{Real-world networks}
In this section we discuss the results obtained by analyzing different graphs describing real-world networks datasets.
Summary statistics of these networks are reported in Table \ref{centrality:tab:datasets-community}.

In particular, Datasets 1--5 represent the undirected networks of coauthors of article appeared in Arxiv\footnote{Arxiv (http://arxiv.org/) is an online archive for scientific preprints in the fields of Mathematics, Physics and Computer Science, amongst others.}, as of April 2003 \cite{leskovec2006sampling}, in the field of, respectively, General Relativity and Quantum Cosmology -- \emph{CA-GrQc}, High Energy Physics (Theory) -- \emph{CA-HepTh}, High Energy Physics (Phenomenology) -- \emph{CA-HepPh}, Astro Physics -- \emph{CA-AstroPh}, and Condensed Matter Physics -- \emph{CA-CondMat}.

Dataset 6 describes a small sample of the Facebook network, representing its directed friendship graph \cite{viswanath2009evolution}.
Finally, dataset 7 represent a large sample of the Facebook network collected by Gjoka et al. \cite{gjoka2010walking}.

This experiment has been designed to quantitatively evaluate the performance of our strategy in real-world applications.
To configure the ERW-Kpath, the values of $\rho$ and $\beta$ have been tuned as previously suggested.
In addition, the value of the maximum length of the self-avoiding random walks has been set equal to $\kappa = 20$.

The results are measured by means of the value of \emph{network modularity} (formally defined by Equation \ref{eqn:qmodexp}) obtained by CONCLUDE, compared against those attained by using three different techniques: \emph{(i)} the already presented \emph{Louvain method} (LM), \emph{(ii)} COPRA \cite{gregory2007algorithm}, which is a fast clustering detection algorithm based on the principle of label propagation \cite{raghavan2007near}, and, finally {\em (iii)} OSLOM \cite{lancichinetti2011finding}, a local optimization algorithm able to finding statistically significant clusters.

Prior to presenting the results of our tests, we briefly describe the main features of COPRA and OSLOM.

COPRA (Community Overlap PRopagation Algorithm) relies on a {\em label propagation strategy} proposed for the first time by Raghavan \emph{et al.} in \cite{raghavan2007near}.
The algorithm works in three stages: {\em (i)} Initially, each vertex $v$ is labeled with a set of pairs
$\langle c,b \rangle$, being $c$ a community identifier and $b$ ({\em belonging coefficient}) a
coefficient indicating the strength of the membership of $v$ to the community $c$; belonging
coefficients are {\em normalized} so that the sum of all the belonging coefficients associated
with $v$ is equal to 1. Initially, the community associated with a vertex coincide with the vertex
itself and the belonging coefficient is 1. {\em (ii)} Then, repeatedly, $v$ updates its label so that
the set of community identifiers associated with $v$ is put equal to the union of the
community identifiers associated with the neighbors of $v$; after that, the belonging coefficients
are updated according to the following formula
$$
b_i(c,v) = \frac{\sum_{w \in N(v)}b_{i-1}(c,v)}{|N(v)|}
$$

being $N(v)$ the set of neighbors of $v$ and $b_{i}(c,v)$ the belonging coefficient associated with
$v$ at the $i$-th iteration. At each iteration, all the pairs in the label of $v$ having a
belonging coefficient less than a threshold are filtered out; in such a case the membership of $v$
to one of the deleted communities is considered not strong enough. It is possible that all the
pairs in a vertex label have a belonging coefficient less than the threshold. In such a case, COPRA
retains only the pair that has the greatest belonging coefficient and deletes all the others.
Finally, if more than one pair has the same maximum belonging coefficient, below the threshold,
COPRA selects at random one of them and this makes the algorithm non-deterministic. After deleting
pairs from the vertex label, the belonging coefficients of each remaining pair are re-normalized so
that they sum to 1. A stopping criterion ensures COPRA ends after a finite number of steps. In such
a case, the set of community identifiers associated with $v$ identify the communities to which $v$
belongs to.

OSLOM (Order Statistics Local Optimization Method) is a {\em multi-purpose} technique that aims at managing directed and undirected graphs. OSLOM is also able to detect overlapping communities and to build
hierarchies of clusters.

The strategy of the algorithm to discover clusters in a graph $G$ is as follows: at the beginning a vertex $i$ is selected at random and it forms the first cluster $C = \{i\}$. After that, the $q$ {\em most
statistically significant} vertices in $G$ are identified and added to $C$. Here $q$ is a random
number and the significance of a vertex $v$ is a parameter indicating the likelihood that $v$ can
be inserted in $C$. To formally define the statistical significance, OSLOM considers a {\em random
null model}, i.e., a class of networks without community structure. A network $G^{'}$ in the random
null model is generated by first copying all the vertices of $G$ in $G^{'}$. After that, multiple
pair of edges in $G^{'}$ are selected at random and an edge is drawn between them. Due to this
procedure, given a vertex $v$ in $G$, there will exist a vertex $v^{'}$ in $G^{'}$ corresponding to
$w$. Analogously, given a subgraph $C$ in $G$, there will be a subgraph $C^{'}$ in $G^{'}$
corresponding to $C$ such that each vertex in $C^{'}$ corresponds to a vertex in $C$. The null model
{\em is expected} not to have a community structure and, therefore, it can be used as a
benchmark to understand if a subgraph $C$ in $G$ is a community and to define the statistical
significance of a vertex $v$ to $C$. In particular, we count the number $l_1$ of vertices linking
$v$ with vertices in $G$; after that, we consider the vertex $v^{'}$ corresponding to $v$ in
$G^{'}$ and we count the number $l_2$ of edges linking $v^{'}$ with vertices residing in $C^{'}$.
If $l_1 > l_2$ we guess that $v$ is significant to $C$ (and can be included in it.)

A community $C$ can be associated with a {\em score} representing its {\em quality}; the score of a
cluster $C$ indicates to what extent $C$ contains vertices which have a high statistical
significance with it. The main idea of OSLOM is to {\em progressively} add and remove vertices
within $C$ so that to improve its score; this procedure is called {\em clean-up}.

The whole process introduced above is repeated several times starting from different nodes in
order to explore different regions of $G$. This yields a final set of clusters that may overlap.

The outcome of our experimental tests are reported in Table \ref{centrality:tab:datasets-community}.
From the analysis of results reported in Table \ref{centrality:tab:datasets-community}, we can draw some consideration about the performance of the proposed clustering method.

Considering these real-world scenarios, CONCLUDE outperforms all its competitors in terms of attained values of \emph{network modularity}.
In general, results provided by CONCLUDE are better than those provided by the \emph{Louvain method} in average of 5\%-15\%, pushing the improvement up to 25\% in the case of COPRA and OSLOM.
This advantage can be explained considering two different motivations: \emph{(i)} our strategy aims at the maximization of the network modularity of a weighted network, producing weights according to an intrinsic rationale, driven by the ERW-Kpath algorithm, while neither the \emph{Louvain method} nor COPRA or OSLOM (and, in general, most of the state-of-the-art clustering algorithms \cite{fortunato2010community}) are able to produce a weighting for any unweighted network.

In addition and equally important, our strategy relies both on local and global information, an aspect which makes CONCLUDE what in recent literature is called a \emph{glocal} optimization algorithm.
In fact, the first step of our strategy exploits information on a long-range, carrying out a \emph{random walker} that visits, starting multiple times from each node, not only the neighborhood, but also those regions of the graph far from the origin of the walk.
This global information is exploited within the second step, the computation of the distance among all pairs of vertices.
Finally, local information is exploited by the modularity optimization strategy inspired by the \emph{Louvain method} itself.
The final result is a general improvement of the performance of the clustering procedure of a non-negligible factor, which comes at almost no cost (in fact, the quality of the partition is proven very good by considering the values of NMI provided in the previous experiment.)

\begin{table}[!ht]
	\centering
	\begin{tabular}{|c l c c c c c c|}
		\hline \hline
		N. &		Network 			&	No. vertices	\	&	\ No. edges	\	& \ CONCLUDE \ & 	\ LM \ & \ COPRA \ & \ OSLOM \ \\	
		\hline \hline
		1	&	CA-GrQc			&	5,242		&	28,980		& \textbf{0.883}	& 0.860	& 0.761	&	0.696\\
		2	&	CA-HepTh			&	9,877		&	51,971		& \textbf{0.806}	& 0.772	& 0.768	&	0.653\\
		3	&	CA-HepPh			&	12,008		&	237,010		& \textbf{0.760}	& 0.656	& 0.754	&	0.675\\
		4	&	CA-AstroPh		&	18,772		&	396,160		& \textbf{0.663} & 0.627	& 0.577	&	0.596\\
		5	&	CA-CondMat		&	23,133		&	186,932		& \textbf{0.768}	& 0.731	& 0.616	&	0.692\\
		6	&	Facebook-links	&	63,731		&	1,545,684	& 0.664 & 0.626	& \textbf{0.726}	&	0.391\\
		7	&	SocialGraph		&	613,497		&	2,045,030	& \textbf{0.912} & 0.891	& 0.197 & 	0.456\\
		\hline \hline
	\end{tabular}
	\caption{Values of \emph{network modularity} provided by CONCLUDE in the context of cluster detection from different \emph{real-world network} datasets. Our algorithm outperforms three of the state-of-the-art solutions, e.g., Louvain method, COPRA and OSLOM.}
	\label{centrality:tab:datasets-community}
\end{table}

\subsection{ERW-Kpath and $\kappa$-path edge centrality} \label{sec:kpath-effect}
In addition to discuss the performance of the clustering method, we here describe some empirical evidence of the $\kappa$-path edge centrality measure as approximated by means of the ERW-Kpath algorithm, which is adopted by CONCLUDE to rank edges.
This is instrumental to understanding the functioning of the \emph{glocal} optimization.

In particular, in the following we report an experiment aimed at discovering how different values of $\kappa$ impact on the final edge centrality.
To this purpose, we produce a probability distribution of $\kappa$-path values obtained varying the setting for $\kappa$ to understand its general behavior.

In detail, we consider the datasets presented in Table \ref{centrality:tab:datasets-community} separately and apply the ERW-Kpath algorithm varying the $\kappa$-path length as $\kappa = 5, 10, 20$.
After that, for a fixed value of $\kappa$-path edge centrality $\overline{L}$, we compute the probability $Pr(\overline{L})$ of finding an edge with such a centrality value.
The corresponding results are plotted in Figure 1 for the top four largest datasets (namely, \emph{CA-HepPh}, \emph{CA-AstroPh}, \emph{CA-CondMat} and \emph{Facebook-links}.)
To show the scaling behavior of the distributions, for each plot we adopt a logarithmic scale.

The analysis of these plots highlights some relevant facts.
First of all, a heavy-tailed distribution in the edge centrality values emerges\footnote{In a \emph{log-log} scale, a distribution that resembles a straight line depicts a scale-free behavior.} in presence of all different values of $\kappa$.
In other words, if we use different values of $\kappa$, the centrality indexes may change; however, as emerges from the plots, the curves representing $\kappa$-path centrality values resemble {\em parallel} lines.
This implies that, for a fixed value of $\kappa$, say $\kappa = 5$, an edge $\overline{e}$ will have a particular centrality score.
But, if $\kappa$ is increased from 5 to 10 and, then, from 10 to 20, the centrality of $\overline{e}$ will be increased always by a constant factor.

This aspect reflects the ability of the ERW-Kpath algorithm to identify those edges which are in fact central in the structure of the network, and rewarding them with high weights.
Intuitively, also those edges which are less relevant will be still awarded, even if for a smaller number of times, which will lead to lower values of centrality (and, in the end, to the heavy-tailed distribution.)
These weights, which are computed according to a global rule (that is, discovering by means of a random walker those edges which are more likely to be traversed during a process of spreading information on the given network) are subsequently exploited to compute overall distances among pairs of vertices.
Eventually, the distances computed on the base of global information, will be exploited to identify the network clusters according to local optima achieved by the modularity optimization greedy strategy.
This summarizes the \emph{glocal} optimization nature of CONCLUDE.

\begin{figure}[!ht]
	\includegraphics[width=.5\columnwidth]{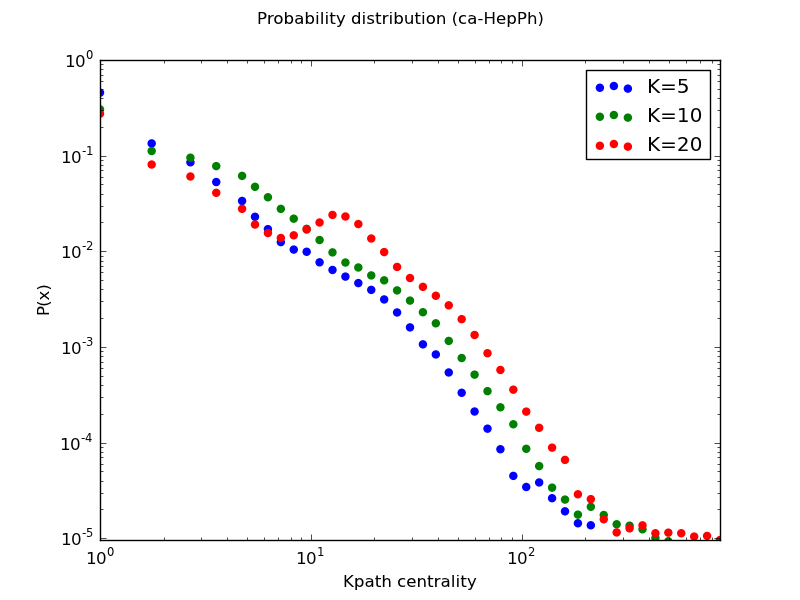}	
	\includegraphics[width=.5\columnwidth]{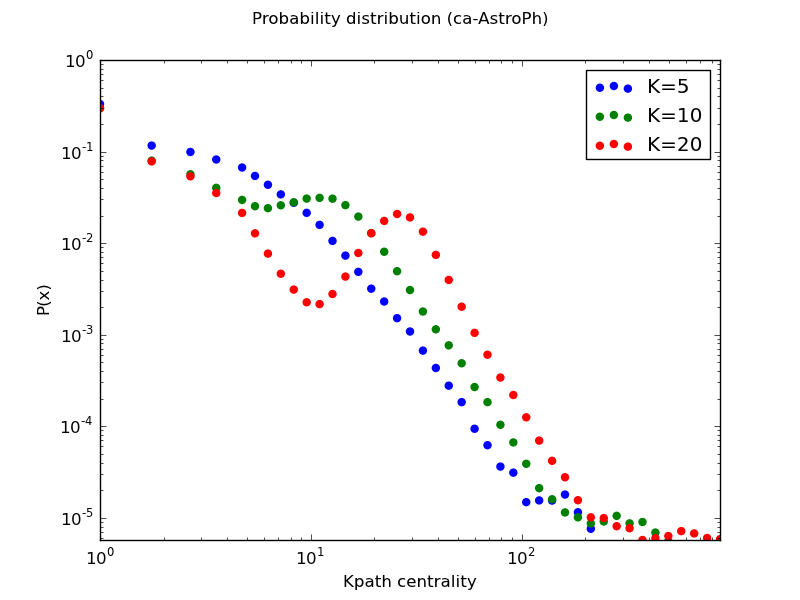}
	\includegraphics[width=.5\columnwidth]{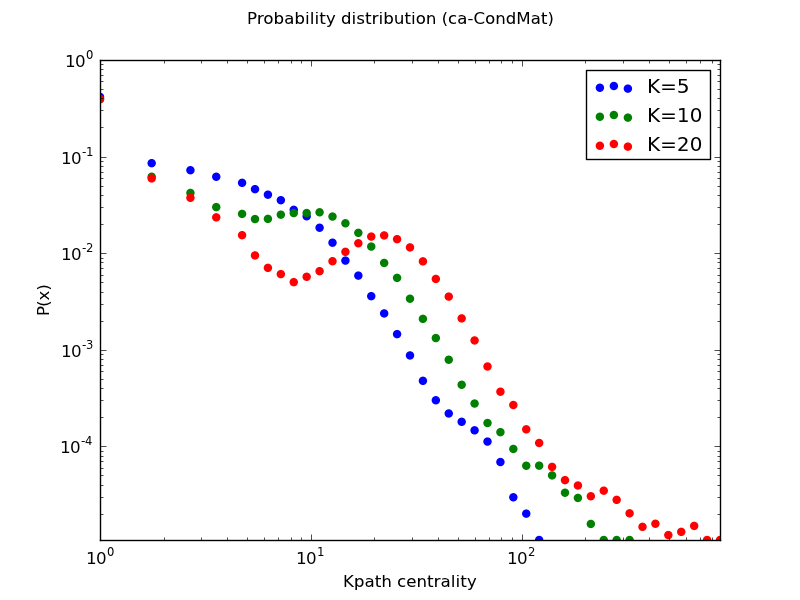}
	\includegraphics[width=.5\columnwidth]{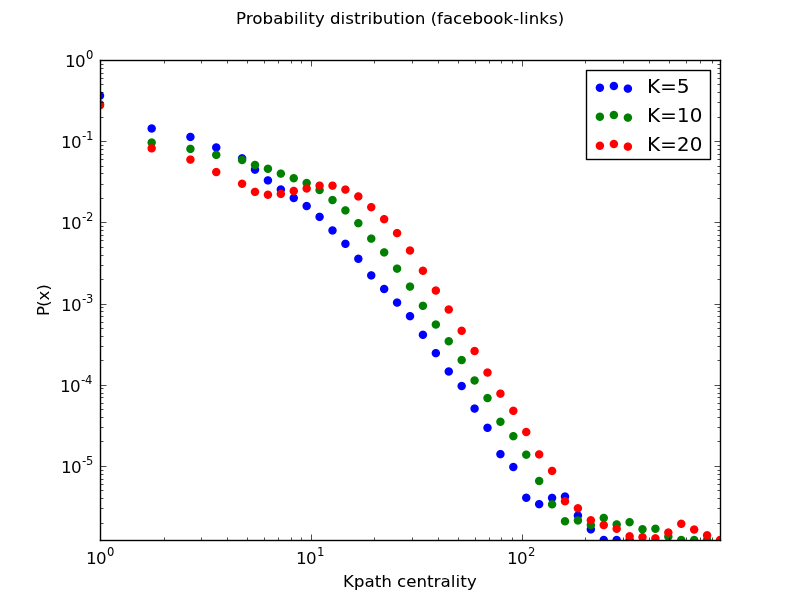}
	\includegraphics[width=.5\columnwidth]{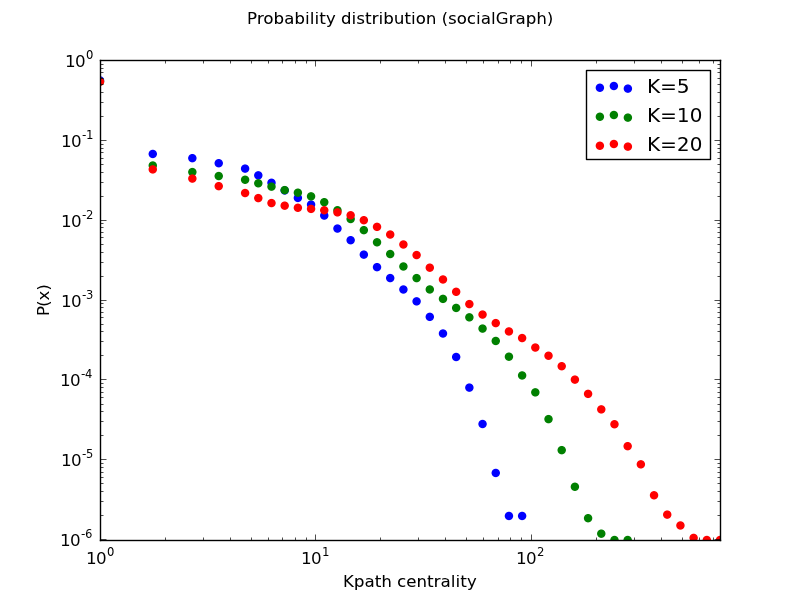}
	\label{fig:k-path-centrality}
	\caption{The probability distribution of $\kappa$-path edge centrality, computed according to different configurations of $\kappa=5,10,20$ for the five largest networks considered in this article: \emph{CA-HepPh}, \emph{CA-AstroPh}, \emph{CA-CondMat}, \emph{Facebook-links} and \emph{SocialGraph}.}
\end{figure}


\section{Conclusions} \label{sec:conclusions}

This article presents CONCLUDE, an efficient method for detecting clusters in complex networks which is proven to work well in different domains.
An early implementation of this algorithm has been already released and its strengths and performance might be assessed independently by other authors.

Our ongoing research efforts focus on adopting CONCLUDE in several contexts.
We are investigating:
\emph{(i)} the emergence of a community structure in large online social networks such as Facebook \cite{ijsnm2011,ferrara2011large};

\emph{(i)} the assessment of sociological conjectures that involve finding clusters according to importance of edges, for example the \emph{strength of the weak ties} theory \cite{granovetter1973strength}, and
\emph{(iii)} enhancing the performance of different state-of-the-art clustering algorithms (such as COPRA \cite{gregory2007algorithm} or OSLOM \cite{lancichinetti2011finding}) by pre-processing networks by means of the random walk based measure of centrality like the $\kappa$-path edge centrality \cite{infsci2012}.

Finally, a long-term research evaluation of our method is planned, in order to cover different domains of application: for example, the application of CONCLUDE could be promising in the context of Neuroinformatics, applied to the \emph{connectome} (i.e., the human brain functional network) \cite{meunier2009hierarchical} or \emph{Bioinformatics,} to detect protein complexes in protein-interaction networks \cite{nepusz2012detecting}.


Further extensions of CONCLUDE will be designed to face additional scientific challenges, such as the possibility of discovering overlapping clusters.
So far, our algorithm is able to produce a strong partition of the network, but does not allow partitions to overlap each other.
This feature will be instrumental in some contexts in which is meaningful that nodes contemporary belong to different clusters (for example, in the case of social networks, users might be part of different communities at the same time.)

\bibliographystyle{abbrv}

\bibliography{jcss-2012-bib}

\end{document}